\documentclass[reprint,amsmath,amssymb,aps,prl]{revtex4-2}
\usepackage[top=2cm, bottom=2cm, left=2cm, right=2cm]{geometry}
\usepackage{graphicx}
\usepackage{amsmath}
\usepackage{xcolor}
\usepackage{float}
\usepackage{bm}
\usepackage[normalem]{ulem}

\usepackage{amsfonts}

\usepackage{array}

\newcommand{\bu}{\boldsymbol{u}}
\newcommand{\bdot}{\boldsymbol{\cdot}}
\newcommand{\bomega}{\boldsymbol{\omega}}

\newcommand{\bez}{\boldsymbol{\widehat{e}_z}}
\newcommand{\bk}{\boldsymbol{k}}
\newcommand{\bx}{\boldsymbol{x}}
\newcommand{\mybullet}{\boldsymbol{\cdot}}
\newcommand{\matB}{\mathbb{A}}
\newcommand{\id}{\mathbb{I}}

\newcommand{\background}[1]{\overline{ #1}}

\begin{document}

\title{{Growth of helicity in salt-finger convection} in the two-dimensional three-component limit}

\author{Smiron Varghese}
\affiliation{CNRS, École Centrale de Lyon, INSA Lyon, Université Lyon 1, LMFA, UMR5509, 69134 Écully, France}

\author{ Benjamin Miquel}
\email[]{benjamin.miquel@cnrs.fr} 
\affiliation{CNRS, École Centrale de Lyon, INSA Lyon, Université Lyon 1, LMFA, UMR5509, 69134 Écully, France}

\author{Wouter J.T. Bos}
\affiliation{CNRS, École Centrale de Lyon, INSA Lyon, Université Lyon 1, LMFA, UMR5509, 69134 Écully, France}

\date{\today}

\begin{abstract}
We present an analytical investigation of the global helicity budget associated with the salt-fingering instability  within the two-dimensional, three-component framework. Our analysis shows that in the region of parameter space corresponding to salt-fingering, helicity {amplification} occurs when the Rayleigh ratio—quantifying the relative strength of salinity to temperature gradients—exceeds a critical threshold. This finding provides a possible theoretical explanation for the spontaneous emergence of helicity, observed in recent numerical simulations.
\end{abstract}

\maketitle

\paragraph{Introduction.}
Helicity is defined as 
\begin{equation}
H=\left\langle\boldsymbol{ u \cdot \omega}\right\rangle\, ,
\end{equation}
where $\bu$ is the velocity and $\bomega\equiv \boldsymbol{ \nabla} \times \bu$ the vorticity. The brackets indicate a volume average. Helicity characterizes, for a given volume, the predominance of helical motion with a given chirality. If in a given velocity field helical motion ($\boldsymbol{ u \cdot \omega}>0$) is equally probable as anti-helical motion ($\boldsymbol{ u \cdot  \omega}<0$), $H$ is expected to fluctuate around zero. The observation of a finite value of $H$ is therefore associated with a kind of symmetry breaking.

As first pointed out by Moreau~\cite{moreau1961constantes} and Moffatt~\cite{moffatt1969degree}, helicity is an invariant of the incompressible Euler equations. The study of helicity subsequently gained interest in the geophysical and astrophysical community, since it might play an important role in the generation of planetary magnetic fields \cite{MoffattBook,Moffatt1992}.
In this context, the origin of helicity is thus an important question.

Recent numerical experiments in periodic geometry have shown that helicity can be amplified through the interplay of unstable stratification and a strong magnetic field \cite{agoua2021spontaneous}. In particular, helicity growth was observed once the flow approached a quasi-two-dimensional three-component (2D3C) state characterized by weak variations  along the vertical direction. 
 
2D3C flows are idealized representations of a number of flows in the presence of body-forces, for instance rapidly rotating flows \cite{gallet2015exact} and conducting fluids under the influence of magnetic fields \cite{gallet2015exact2,favier2010two}. In this limit the helicity degenerates to a correlation between the passively advected vertical velocity $w$ and the vertical vorticity $\omega_z$ \cite{biferale2017two,yin2024influence}. The {amplification} of helicity observed in \cite{agoua2021spontaneous} was suggested to be associated with large scale spiraling upwards and downwards motion in the 2D3C limit.
 
More recently, numerical studies have shown that the double-diffusive instability (sometimes dubbed salt-finger convection) can also generate helicity in a stably stratified flow~\cite{fraser2025helical}. 
In the present study, we analytically investigate helicity {amplification} by double-diffusion in the early stages of the growth of the perturbation. Specifically, we focus on linear dynamics pertaining to small perturbations in the 2D3C limit. We derive a simple model for linear helicity {amplification}.
 
The outcome is that, for the parameters where a stable stratification can become unstable through salt-fingering, the 2D3C mechanism underlying helicity {amplification} previously observed in \cite{agoua2021spontaneous} is present. We emphasize, however, that by contrast with the unstably stratified case which is inviscid by essence, the salt-fingering analysis relies on retaining all terms entering the linear dynamics, including the diffusive terms, due to the intrinsically diffusive nature of the phenomenon.

{Importantly, even in the absence of solid boundaries, this insight provides a possible amplification mechanism for helicity by solely relying on the 2D3C character of the flow.
In the context of stellar and planetary interiors, this generic mechanism favours the amplification of magnetic fields, among other candidates \cite{rincon2019dynamo,tobias2021turbulent}.}

This Letter is organized as follows: first, we write down the governing equations of our system, and we analyze the 2D3C limit. We determine a threshold beyond which linear analysis predicts helicity {amplification}. Next, we compare the numerical values of these bounds with typical values in geo- and astrophysical settings and we discuss our results in the light of recent numerical simulations. Finally, the conclusion summarizes the results.

\paragraph{Formulation.} We consider a base state (denoted by $\background{\cdot}$) that corresponds to a stratified flow initially at rest ($\background{\bu} = 0$), {subjected to vertical gravity $\boldsymbol{g} = -g \bez$ and with both temperature and salinity linearly increasing upwards. 
Precisely, we denote the homogeneous positive temperature gradient  and salt-concentration gradients $\beta_T=\partial_z \background{ T}>0$ and $\beta_S=\partial_z\background{S}>0$, respectively.}
We adopt the Boussinesq approximation and accordingly linearize the temperature and concentration dependence of the density around its reference value $\rho_0=\rho(T_0,S_0)$,  
\begin{equation}
\rho(T,S)=\rho_0\left[1-\alpha_T(T-T_0)+\alpha_S(S-S_0)\right],
\end{equation}
where $\alpha_T,\alpha_S>0$ are thermal expansion and haline contraction coefficients, respectively.
{Thus, the background temperature stabilizes the stratification whereas the background concentration antagonistically tends to destabilize the stratification. However,} we consider the case where the thermal stabilization overcomes the saline destabilisation, resulting in an overall statically stable stratification $\partial \rho / \partial z  < 0$. 
It has been known since the work of Stern \cite{Stommel1956,stern1960salt} that such a density stratification can potentially be destabilized if the diffusivities of salt $\kappa_S$ and temperature $\kappa_T$ are unequal, which is typically the case in sea water where $\kappa_T/\kappa_S\approx 100$. The resulting flow is composed of characteristic structures, named salt-fingers, that consist of vertically elongated slender columns of ascending and descending fluid \cite{schmitt1994,radko2013double,Garaud2018}.

Linear stability can provide hindsight for non-dimensionalization~\cite{stern1960salt,xie2017reduced} by identifying that fastest growing perturbations have a typical lengthscale
\begin{equation}
\label{eq:finger_scale}
d=\left(\frac{\kappa_T\nu}{g\alpha_T \beta_T}\right)^{1/4},
\end{equation}
with $\nu$ the kinematic viscosity. Accordingly, time and velocity are measured in diffusive units, $d^2/\kappa_S$ and $\kappa_S/d$, respectively. Temperature and salinity fluctuations around the initial linear gradient background, are non-dimensionalized using $\beta_T d\kappa_S /\kappa_T$ and $\alpha_T\beta_T d\kappa_S /(\alpha_S\kappa_T)$, respectively. The governing equations read:
\begin{subequations}
\label{sys:governing}
\begin{align}
\frac{1}{Sc}\left(\frac{\partial }{\partial t}+\boldsymbol{u\cdot \nabla } \right)
\bu &=\nabla^2 \bm u-\bm \nabla p+\left(T'  - S'\right)\bez\label{eq:momentum}\,,\\
\frac{1}{Le}\left(\frac{\partial }{\partial t}+\boldsymbol{u\cdot \nabla } \right)T'&=\nabla^2 T'- w\label{eq:temperature}\,,\\
\left(\frac{\partial }{\partial t}+\boldsymbol{u\cdot \nabla } \right)S'&=   \nabla^2 S'-\mathcal{R} w\label{eq:S}\,,\\
\bm \nabla \cdot \bm u&=0\,,
\end{align}
\end{subequations}
where $\boldsymbol{u}=(u,v,w)$ is the dimensionless velocity, and the quantities $T'$ and $S'$ indicate fluctuations of temperature and salinity with respect to the mean-gradient profiles.
In this formulation, the natural control parameters are two diffusivity ratios, the Schmidt number $Sc$ and the Lewis number $Le$, and the Rayleigh ratio $\mathcal{R}$, respectively defined as
\begin{subequations}
\label{control_parameters}
\begin{gather}
Sc=\frac{\nu}{\kappa_S}\,, \label{def:Sc}\\
Le=\frac{\kappa_T}{\kappa_S}\,, \label{def:Le}\\
\mathcal R=\frac{\alpha_S \beta_S}{\kappa_S}\frac{\kappa_T}{\alpha_T \beta_T}\,. \label{def:R}
\end{gather}
\end{subequations}
These three quantities span the parameter space. Alternatively, the diffusivity ratios can be dialed into the Prandtl number $Pr\equiv Sc/Le= \nu/\kappa_T$. In the following subsection we will recall which subspace is relevant for the formation of salt-fingers.
 
\paragraph{Salt-fingering instability.}

Here, we summarize in broad sketches linear stability analysis 
applied to doubly-diffusive convection.
We seek solutions to Eqs.~\eqref{sys:governing} in the form of small amplitude normal modes in space and time
\begin{multline}
\label{eq:perturbation}
\{u,v,w,p,T',S'\}(\bx,t)=\epsilon \{\hat u,\hat v,\hat w,\hat p,\hat T',\hat S'\}\\ \times \exp\left(i\bk\bdot \bm x+\mu t\right)\,,
\end{multline}
where $\bk$ is the wave vector of the perturbation, and $\mu$ is the growth rate. Substituting expression \eqref{eq:perturbation} in \eqref{sys:governing} and collecting leading $O(\epsilon)$ terms yields an algebraic system, where advection of the fluctuations is neglected.
The dispersion relation that implicitly relates the growth rate to the wave vector of the perturbation $\bk$ and all control parameters is obtained by cancelling the associated determinant. Instability {develops} in regions of parameter space where $\Re(\mu)>0$ for some wave vectors $\bk$. It has been shown (e.g.~\cite{bainesJFM69}) that a necessary condition for instability is
\begin{equation}\label{eq:SFcriterion}
1< \mathcal R< Le\,.
\end{equation}
For this reason, hereafter we focus on this parameter range, where the Rayleigh ratio $\mathcal R$ is {bracketed by} unity and the Lewis number.

\paragraph{Helicity amplification.} We now determine whether helicity can be generated by a mechanism similar to the one discussed in Ref.~\cite{agoua2021spontaneous}, where it was observed that in unstable stratification, in the presence of a strong magnetic field, the mean-helicity can grow exponentially, suggesting a linear instability. The close to two-dimensional three-component flow in this limit is, at least visually, not so different from the salt-fingering flow. Therefore, we assume a vertically invariant solution ($\partial_z=0$) {below}. We express the solenoidal velocity as:
\begin{equation}
    \bu = -\boldsymbol{\nabla } \times \psi \, \bez 
    + w \, \bez\,.
\end{equation}
The governing equations for momentum become:
\begin{subequations}
\label{sys:gov_pot}
\begin{gather}
\frac{1}{Sc}\left(\frac{\partial }{\partial t}+J\left[\psi, \mybullet\right] \right) \nabla_\perp^2 \psi = \nabla_\perp^4\psi
\label{eq:psi}\,,\\
\frac{1}{Sc}\left(\frac{\partial }{\partial t}+J\left[\psi, \mybullet\right] \right) w = \nabla_\perp^2 w + T - S
\label{eq:w}\,,
\end{gather}
\end{subequations}
where $J\left[\psi, \mybullet\right] = \partial_x \psi \partial_y - \partial_y \psi \partial_x$ represents horizontal advection and $\nabla_\perp^2=\partial_{xx}+\partial_{yy}$ is the horizontal Laplace operator. Inspired by the analysis in \cite{agoua2021spontaneous}, we focus on the component of the helicity associated with vertically spiraling motion,
\begin{equation}
h_z= w\,\omega_z = w \nabla_\perp^2 \psi\,.
\end{equation}
Combining the governing equations for momentum, temperature, and salinity, we obtain the following dynamical system:
\onecolumngrid
\begin{equation}
\label{eq:dispersion_relation}
    \left(\partial_t + J\left[ \psi, \mybullet \right] \right) \begin{pmatrix}
     h_z \\
     q_T \\ 
     q_S 
    \end{pmatrix} = 
    \matB
  \begin{pmatrix}
     h_z \\
     q_T \\ 
     q_S 
    \end{pmatrix}\,, \quad \text{with}\quad  
    \matB=\begin{pmatrix}
    -2 k_\perp^2 Sc  & Sc & -Sc\\
    -Le  & -k_\perp^2 \left( Sc + Le \right) & 0 \\
    -\mathcal{R} & 0 & -k_\perp^2 (Sc+1)\end{pmatrix}
\end{equation}
where the state vector is formed by the vertical helicity $h_z$, and:
\begin{subequations}
\label{def:q}
\begin{align}
    q_T = T\nabla_\perp^2 \psi \label{def:qt}\,, \\ 
    q_S = S\nabla_\perp^2 \psi \label{def:qS} \,.
\end{align}
\end{subequations}
We define the state vector of volume-averaged quantities (denoted with angle brackets $\left\langle \dots \right \rangle$):
\begin{equation}
 X = \begin{pmatrix}   H_z\\ Q_T\\  Q_S\end{pmatrix} = \begin{pmatrix}
 \left \langle h_z \right \rangle \\ 
 \left \langle q_T \right \rangle \\ 
 \left \langle q_S \right \rangle \end{pmatrix} 
\end{equation}
and assume an exponential growth or decay in time of the form $\exp \left( \gamma t \right)$. Observing that $\left \langle J \left [ \psi, \mybullet \right] \right \rangle =0$ the volume-averaged quantities obey the eigenvalue problem:
\begin{equation}
\label{eq:ev}
    \gamma X = \matB X
\end{equation} 
We first note that the diagonal terms of $\matB$, all proportional to $k_\perp^2$, are of viscous or diffusive nature. If we neglect these terms, in the spirit of the work in~\cite{agoua2021spontaneous}, we readily obtain the linear growth rate
\begin{equation}
\gamma^2=Sc~(\mathcal R -Le)\,.
\end{equation}
Thus, in this inviscid approximation, a necessary condition for  the linear growth of helicity is:
\begin{equation}
\mathcal R>Le .
\end{equation}
This is in contradiction with the salt finger instability criterion given in Eq.~\eqref{eq:SFcriterion}. However, due to the diffusive nature of the instability, it is natural that diffusive mechanisms have a role to play in the putative linear helicity {amplification}, in contrast with the scenario reported in~\cite{agoua2021spontaneous}. For this reason, we abandon our inviscid assumption and now turn to computing the exact spectrum of $\matB$ where all the diffusive terms have been retained. Cancelling the determinant $|\gamma \id - \matB|$ yields:
\begin{equation}
 \left(\gamma + 2k_\perp^2 Sc\right)    
 \left(\gamma +  k_\perp^2 [Sc+Le]\right)    
 \left(\gamma +  k_\perp^2 [Sc+1]\right) 
  + Sc Le \left(\gamma + k_\perp^2 [Sc+1]\right)
  - \mathcal{R}Sc  \left(\gamma + k_\perp^2 [Sc+Le]\right)    = 0 \,,
\end{equation}
which is of the form $P_3(\gamma)=0$. Expanding the third order polynomial $P_3$, one obtains:
\begin{multline}
\label{eq:p3}
    \gamma^3 + \gamma^2 k^2_\perp \left(4 Sc + Le + 1 \right)
              + \gamma  k^4_\perp\left(5 Sc^2 + 2 Sc Le + 3Sc + Le\right)
              + \gamma Sc \left( Le - \mathcal{R}\right) \\ 
	      + k^2_\perp Sc \left( Sc \left[Le-\mathcal{R}\right] - Le \left[\mathcal{R}-1\right]\right) + 2 Sc \left( Sc + Le \right) \left( Sc + 1 \right) k_\perp^6 = 0\,.
\end{multline}
In the salt-finger regime characterized by \eqref{eq:SFcriterion}, it is straightforward that all non-constant coefficients of $P_3$ [the upper line of~\eqref{eq:p3}] are positive, upon inspection. According to Descartes' rule of signs~\cite{descartes}, one positive real root can exist if, and only if, the constant coefficient of \eqref{eq:p3} is negative. For a given horizontal wave number $k_\perp$,
this is the case for  $\mathcal{R}>\widetilde{\mathcal{R}}_H(k_\perp)$ where
\begin{equation}\label{eq:Rcrit}
	\widetilde{\mathcal{R}}_H \equiv \frac{ Sc + 1 }{Sc + Le } Le  + 2 \left( Sc + 1 \right) k_\perp^4 \,.
\end{equation}
This expression is minimized for $k_\perp \to 0$, so that the onset for helicity amplification $\mathcal{R}_H$ is equivalently defined with:
\begin{equation}\label{eq:Rcrit}
	\mathcal{R}_H \equiv \min_{k_\perp} \widetilde{\mathcal{R}}_H \left( k_\perp \right) = \frac{Sc+1}{Sc+Le} Le = \frac{Sc+1}{Pr+1} = \frac{PrLe+1}{Pr+1} \,.
\end{equation}

Consistently with our assumption that~\eqref{eq:SFcriterion} holds, we observe that the threshold value for linear helicity {amplification} $\mathcal{R}_H$ is found within the salt-finger instability interval~\eqref{eq:SFcriterion} for $Le>1$. The parameter range for helicity amplification is finally obtained as the intersection of~\eqref{eq:SFcriterion} and~\eqref{eq:Rcrit}, namely:
\begin{equation}
    \label{eq:helicity_range}
    \mathcal{R}_H<\mathcal{R}<Le\,.
\end{equation}

\begin{table*}[]
    \centering
\begin{center}
\begin{tabular}{ c | c c c c | c}
\hline
Fluid &~~ $Pr\equiv \nu /\kappa_T$~~ & ~~ $Sc\equiv \nu / \kappa_S $ ~~ &~~ $Le\equiv \kappa_T/\kappa_S$ ~~ & ~$\mathcal{R}_H$ ~& ~ helicity amplification range\\ 
\hline
Earth's  oceans (water) & $7$  &  $700$ & $100$ & $87.625$ ~ & $87.625<\mathcal{R}<100$ \\
Planetary cores (Liquid metal)~ & $10^{-2}$ & $10^{2}$ & $10^4$ & $100$ ~& $10^2\lesssim \mathcal{R}<10^4$ \\
Stellar interiors 1 (Plasma) & $10^{-6}$ & $10^1$ & $10^{7}$ & $10$~& $10\lesssim \mathcal{R}<10^{7}$\\
Stellar interiors 2 (Plasma) & $10^{-6}$ & $10^{-3}$ & $10^{3}$ & $1$~& $1\lesssim \mathcal{R}<10^3$\\
\hline
\end{tabular}
\end{center}
    \caption{Values of dimensionless parameters (Prandtl $Pr$, Schmidt $Sc$, Lewis $Le$ numbers) for oceans~\cite{schmitt1994}, planetary cores~\cite{Dormy2025} and stellar interiors~\cite{Garaud2018}. For each system, we compute the corresponding critical value for helicity {amplification} $\mathcal{R}_H$ as predicted by~\eqref{eq:Rcrit} and explicitly indicate the range $\mathcal{R}_H<\mathcal{R}<Le$ over which helicity is linearly amplified, for completeness. Except for the sea-water values, all values are order of magnitude estimates. Due to the large variability in the value of the Schmidt number in stars, two limit cases are considered.}
    \label{tab:Rcrit}
\end{table*}

\begin{figure}
    \centering
    \includegraphics[width=0.495 \textwidth]{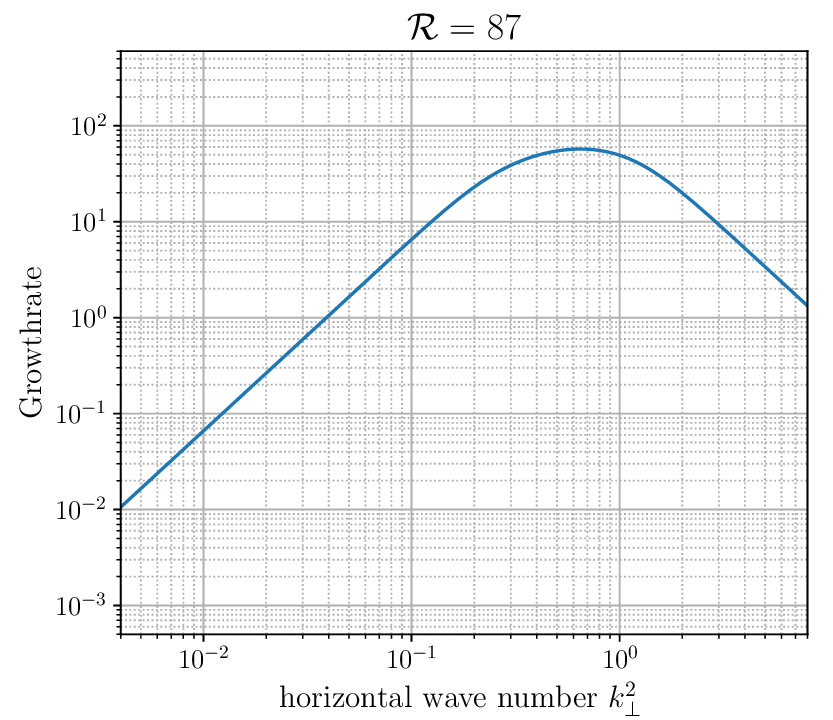}
    \includegraphics[width=0.495 \textwidth]{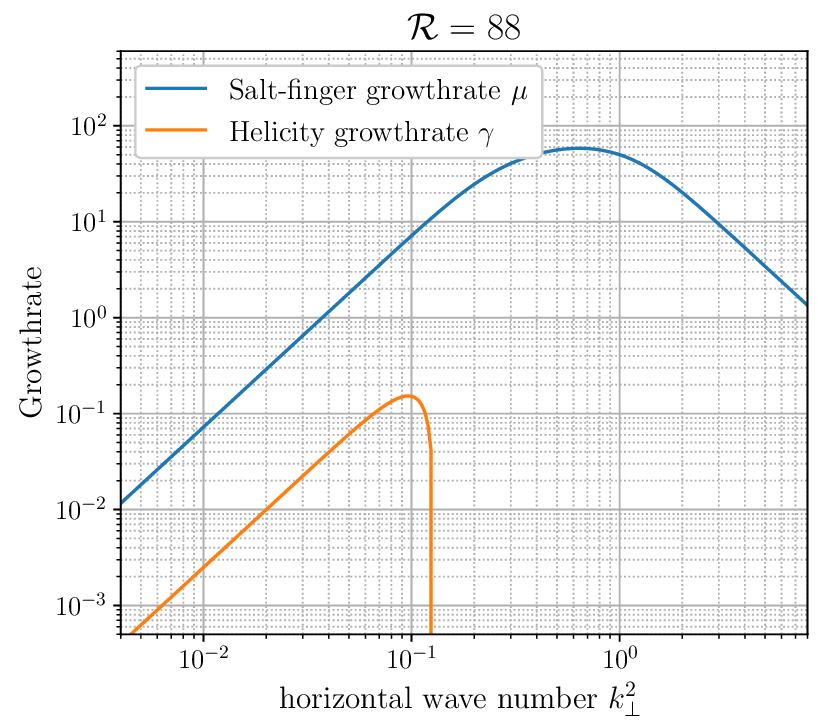}
    \caption{Growth rates of the salt-fingers $\mu$ and of helicity $\gamma$ computed for sea water values of parameters $Sc=700$ and $Le=100$ as a function of the horizontal wave number squared. We compare $\mathcal{R}=87$ (left panel) and $\mathcal{R}=88$ (right panel), which conveniently bracket the expected threshold value $\mathcal{R}_H=87.625$. As expected, we observe that the growth rate of salt-fingers (blue lines) varies imperceptibly between the two case. This is in strong contrast with the helicity growth rate: above onset, $\gamma>0$ (orange line, right panel), whereas $\gamma<0$ below onset and therefore is absent from the semi-log representation of the left panel.}
    \label{fig:helicity_generation}
\end{figure}
\newpage
\twocolumngrid
\paragraph{Bounds for geo- and astrophysical flows.}
Below we comment on characteristic values that pertain to three classes of flows: water oceans, liquid metal planetary cores, and stellar cores (see Table~\ref{tab:Rcrit}). {We also compare our findings to the numerical experiments of \cite{fraser2025helical}.}
\paragraph{a. Water oceans.} For sea water on Earth, adopting $Sc=700$ and $Le=100$, we find a threshold value of $\mathcal{R}_H=87.625$. Numerically, we confirm on figure~\ref{fig:helicity_generation}, where salt finger and helicity growth-rates are computed for different values of $\mathcal{R}$, that growing helical modes are found for $\mathcal{R}>\mathcal{R}_H$. 
Helicity amplification occurs over the range defined by equation~\eqref{eq:helicity_range}, which for sea water translates to  $87.625<\mathcal{R}<100$.
\paragraph{b. Planetary cores---the case of liquid metal.} Liquid metals that compose planetary cores are characterized by the diffusivity hierarchy $\kappa_T \gg \nu \gg \kappa_S$, or equivalently $Le \gg Sc \gg 1 \gg Pr$. In addition to planetary cores, this limit is also relevant to some stars (``Stellar interiors 1'' in table~\ref{tab:Rcrit}). In this geophysically relevant case, 
expression \eqref{eq:Rcrit} simplifies so that the critical density ratio is approximately $\mathcal{R}_H \approx Sc$. Thus, the range for linear helicity amplification by 2D3C salt-fingering flows becomes
\begin{equation}\label{eq:range_planets}
    Sc \lesssim \mathcal{R} < Le \,.
\end{equation}

We note that, in this regime, the ratio between the upper and the lower bound of the helicity amplification interval simplifies to $Le/Sc=1/Pr \gg 1$. This interval is thus much wider for planetary cores than for the case of sea water discussed in the previous paragraph,
underlining the plausibility of helicity amplification
through the present laminar mechanism in a large part of the parameter space, provided the flow is close enough to the 2D3C limit. 

\paragraph{c. Stellar cores.} Although some stellar interiors are well described
by the paragraph above, others are characterized by $\kappa_T \gg \kappa_S \gg \nu$, or equivalently $Le \gg 1 \gg Sc \gg Pr$ (labelled ``Stellar interiors 2'' in 
Table~\ref{tab:Rcrit}). The picture clarifies further in this limit, where the helicity amplification onset is approximately $\mathcal{R}_H \approx 1$ so that the range
for helicity amplification almost matches the entire double-diffusion instability 
region:
\begin{equation}\label{eq:range_stars}
    1 \lesssim \mathcal{R} < Le \,.
\end{equation}
Therefore, we conclude that in both planetary and stellar cores, helicity amplification is possible through the present laminar mechanism over a significant 
fraction of the parameter space region where salt-fingering occurs, provided the flow is close enough to the 2D3C limit. 

\paragraph{Comparison to numerical experiments.}
In the direct numerical simulations of \cite{fraser2025helical}, salt-finger convection was studied in elongated boxes in the fully developed (nonlinear) regime. In the statistically steady state resulting from the saturated salt-fingering instability, a helical mean-flow was observed in all their simulations. In the three simulations reported in their system, the {diffusivity parameters were fixed to $Pr=5$, $Le=20$, $Sc=100$ while the Rayleigh ratio was chosen in $\mathcal{R}=1.01, 1.1, 2$.} For these parameter values, our analysis predicts an interval for linear amplification $16.8<\mathcal{R}<20$. {Therefore, helicity amplification occurs in a wider interval in the nonlinear DNS of~\cite{fraser2025helical} than predicted by our linear analysis. This observation is consistent with the fact that nonlinear dynamics allows for additional scenarios for helicity amplification compared to linear dynamics, whose predictions merely yield a sufficient (but not necessary) condition for helicity amplification.
 
Regarding the saturation of linear amplification of helicity, we refer to \cite{garaud20152d}, where it is discussed how the fastest-growing vertically invariant salt-fingers undergo a secondary 3D shearing instability (dependent on the domain size), breaking the 2D3C character of the flow and entering the fully nonlinear regime. We remark that the saturation of helicity is a direct consequence of the saturation of the fingers as the Cauchy-Schwarz inequality readily provides an upper bound, $\langle w\omega_z\rangle\leq\langle w^2\rangle^{1/2}\langle \omega_z^2\rangle^{1/2}$. Incidentally, for unstably stratified flows considered in reference \cite{agoua2021spontaneous}, it was observed that the largest structures of the flow approached this limit of maximal helicity.

\paragraph{Conclusion}
The above analysis shows that in the range where salt-fingering occurs, helicity can be generated if the flow is close-enough to the 2D3C limit. This analysis sheds light on the recent work by Fraser and coworkers~\cite{fraser2025helical}  which showed the generation of helicity in a salt-fingering flow and \cite{agoua2021spontaneous} for unstable stratification. From our analysis it follows that a common helicity amplification mechanism can manifest in both cases, despite their differences.

Indeed, helicity amplification exists in both configurations, provided the flow approaches the 2D3C limit to a good approximation. The exact driving force, however, differs. In \cite{agoua2021spontaneous} helicity {amplification} results from the unstable density stratification and is thus intrinsically an inviscid process. By contrast, it is the double-diffusive stability which drives the dynamics responsible for helicity {amplification} in \cite{fraser2025helical}.
An interesting perspective is to further investigate this result for the case of the liquid metal cores of celestial bodies, and in particular the growth of a magnetic field via the dynamo effect. Indeed, we show in Table \ref{tab:Rcrit} that the present results can be extremely relevant to explain global helicity {amplification}. Since the present results suggest that, in planets and stars, helicity can be generated by the double diffusive instability for $Sc<\mathcal{R}<Le$, helical salt-fingering might be considered as a mechanism to excite and sustain planetary dynamos.

\begin{acknowledgments}
This research was funded, in whole or in part, by Agence Nationale de la Recherche (Grant ANR-23-CE30-0016-01). 
We thankfully acknowledge Adrian Fraser for kindly pointing out an omission in an earlier version.
\end{acknowledgments}

%
\bibliographystyle{unsrt}
\bibliography{biblio}

@book{radko2013double,
  title={Double-diffusive convection},
  author={Radko, Timour},
  year={2013},
  publisher={Cambridge University Press}
}

@book{descartes,
title={Discours de la m\'ethode: la G\'eom\'etrie},
year={1637},
author={Ren\'e Descartes}
}

@article{schmitt1994,
  title = {Double Diffusion in Oceanography},
  volume = {26},
  ISSN = {1545-4479},
  number = {1},
  journal = {Annu. Rev. Fluid Mech.},
  publisher = {Annual Reviews},
  author = {Schmitt,  R W},
  year = {1994},
  pages = {255–285}
}

@article{Dormy2025,
  title = {Rapidly Rotating Magnetohydrodynamics and the Geodynamo},
  volume = {57},
  ISSN = {1545-4479},
  url = {http://dx.doi.org/10.1146/annurev-fluid-031224-121649},
  DOI = {10.1146/annurev-fluid-031224-121649},
  number = {1},
  journal = {Annu. Rev. Fluid Mech.},
  publisher = {Annual Reviews},
  author = {Dormy,  Emmanuel},
  year = {2025},
  pages = {335–362}
}

@book{MoffattBook,
  title={Magnetic field generation in electrically conducting fluids},
  author={Moffatt, H.K.},
  year={1978},
  publisher={Cambridge University Press}
}

@article{moreau1961constantes,
  TITLE = {{Constantes d'un {\^i}lot tourbillonnaire en fluide parfait barotrope}},
  AUTHOR = {Moreau, Jean Jacques},
  JOURNAL = {{Comptes rendus hebdomadaires des s{\'e}ances de l'Acad{\'e}mie des sciences}},
  PUBLISHER = {{Gauthier-Villars}},
  VOLUME = {252},
  PAGES = {2810-2812},
  YEAR = {1961}
}

@article{Garaud2018,
  title = {{D}ouble-{D}iffusive {C}onvection at {L}ow {P}randtl {N}umber},
  volume = {50},
  ISSN = {1545-4479},
  DOI = {10.1146/annurev-fluid-122316-045234},
  number = {1},
  journal = {Annu. Rev. Fluid Mech.},
  publisher = {Annual Reviews},
  author = {Garaud,  Pascale},
  year = {2018},
  month = jan,
  pages = {275–298}
}

@article{fraser2025helical, 
title={Spontaneous generation of helical flows by salt fingers}, volume={1020}, DOI={10.1017/jfm.2025.10641}, journal={J. Fluid Mech.}, author={Fraser, Adrian E. and van Kan, Adrian and Knobloch, Edgar and Julien, Keith and Liu, Chang}, year={2025}, pages={R1}}

@article{stern1960salt,
  title={The “salt-fountain” and thermohaline convection},
  author={Stern, Melvin E},
  journal={Tellus},
  volume={12},
  number={2},
  pages={172--175},
  year={1960},
  publisher={Taylor \& Francis}
}

@article{xie2017reduced,
  title={A reduced model for salt-finger convection in the small diffusivity ratio limit},
  author={Xie, Jin-Han and Miquel, Benjamin and Julien, Keith and Knobloch, Edgar},
  journal={Fluids},
  volume={2},
  number={1},
  pages={6},
  year={2017},
  publisher={MDPI}
}

@article{agoua2021spontaneous,
  title={Spontaneous generation and reversal of helicity in anisotropic turbulence},
  author={Agoua, Wesley and Favier, Benjamin and Delache, Alexandre and Briard, Antoine and Bos, Wouter J. T.},
  journal={Phys. Rev. E},
  volume={103},
  number={6},
  pages={L061101},
  year={2021},
  publisher={APS}
}

@article{bainesJFM69,
  title = {On thermohaline convection with linear gradients},
  volume = {37},
  ISSN = {1469-7645},
  url = {http://dx.doi.org/10.1017/S0022112069000553},
  DOI = {10.1017/s0022112069000553},
  number = {2},
  journal = {J. Fluid Mech.},
  publisher = {Cambridge University Press (CUP)},
  author = {Baines,  P. G. and Gill,  A. E.},
  year = {1969},
  month = 6,
  pages = {289–306}
}

@ARTICLE{Moffatt1992,
AUTHOR = "H.K. Moffatt and A. Tsinober",
TITLE = "Helicity in laminar and turbulent flow",
JOURNAL = "Ann. Rev. Fluid Mech.",
VOLUME = "24",
YEAR = "1992",
PAGES = "281"}

@article{moffatt1969degree,
  title={The degree of knottedness of tangled vortex lines},
  author={Moffatt, Henry Keith},
  journal={J. Fluid Mech.},
  volume={35},
  number={1},
  pages={117--129},
  year={1969},
  publisher={Cambridge University Press}
}

@article{Stommel1956,
  title = {An oceanographical curiosity: the perpetual salt fountain},
  volume = {3},
  ISSN = {0146-6313},
  DOI = {10.1016/0146-6313(56)90095-8},
  number = {2},
  journal = {Deep Sea Res.},
  publisher = {Elsevier BV},
  author = {Stommel,  Henry and Arons,  Arnold B. and Blanchard,  Duncan},
  year = {1956},
  pages = {152–153}
}

@article{gallet2015exact,
  title={Exact two-dimensionalization of low-magnetic-Reynolds-number flows subject to a strong magnetic field},
  author={Gallet, Basile and Doering, Charles R},
  journal={J. Fluid Mech.},
  volume={773},
  pages={154--177},
  year={2015},
  publisher={Cambridge University Press}
}

@article{gallet2015exact2,
  title={Exact two-dimensionalization of rapidly rotating large-Reynolds-number flows},
  author={Gallet, Basile},
  journal={J. Fluid Mech.},
  volume={783},
  pages={412--447},
  year={2015},
  publisher={Cambridge University Press}
}

@article{favier2010two,
  title={On the two-dimensionalization of quasistatic magnetohydrodynamic turbulence},
  author={Favier, Benjamin and Godeferd, Fabien S and Cambon, Claude and Delache, Alexandre},
  journal={Phys. Fluids},
  volume={22},
  number={7},
  year={2010},
  publisher={AIP Publishing}
}

@article{biferale2017two,
  title={From two-dimensional to three-dimensional turbulence through two-dimensional three-component flows},
  author={Biferale, Luca and Buzzicotti, M and Linkmann, Moritz},
  journal={Phys. Fluids},
  volume={29},
  number={11},
  year={2017},
  publisher={AIP Publishing}
}

@article{yin2024influence,
  title={Influence of the vorticity-scalar correlation on mixing},
  author={Yin, Xi-Yuan and Agoua, Wesley and Wu, Tong and Bos, Wouter JT},
  journal={Phys. Rev. Fluids},
  volume={9},
  number={10},
  pages={104502},
  year={2024},
  publisher={APS}
}

@article{rincon2019dynamo,
  title={Dynamo theories},
  author={Rincon, Fran{\c{c}}ois},
  journal={J. Plasma Phys.},
  volume={85},
  number={4},
  pages={205850401},
  year={2019},
  publisher={Cambridge University Press}
}

@article{tobias2021turbulent,
  title={The turbulent dynamo},
  author={Tobias, SM},
  journal={J. Fluid Mech.},
  volume={912},
  pages={P1},
  year={2021},
  publisher={Cambridge University Press}
}

@article{garaud20152d,
  title={2{D} or not 2{D}: the effect of dimensionality on the dynamics of fingering convection at low Prandtl number},
  author={Garaud, Pascale and Brummell, Nicholas},
  journal={Astrophys. J.},
  volume={815},
  number={1},
  pages={42},
  year={2015},
  publisher={The American Astronomical Society}
}

\end{document}